\documentclass[twocolumn]{aastex631}
\usepackage{gensymb}
\usepackage{lineno}
\usepackage{emptypage}
\shorttitle{fermi 1544}
\shortauthors{Shao et al.}

\graphicspath{{./}{figures/}}

\begin{document}

\title{Is Fermi 1544-0649 a misaligned blazar? discovering the jet structure with VLBI}

\correspondingauthor{Xiaopeng Cheng}
\email{xcheng0808@gmail.com}

\correspondingauthor{Tam, Pak-Hin Thomas}
\email{tanbxuan@mail.sysu.edu.cn}

\correspondingauthor{Yudong Cui}
\email{cuiyd@mail.sysu.edu.cn}

\correspondingauthor{Lili Yang}
\email{yanglli5@mail.sysu.edu.cn}

\author{Chengyu Shao}
\affiliation{School of Physics and Astronomy, Sun Yat-sen University, Guangzhou 510275, People’s Republic of China}

\author[0000-0003-4407-9868]{Xiaopeng Cheng}
\affiliation{Korea Astronomy and Space Science Institute, 776 Daedeok-daero, Yuseong-gu, Daejeon 34055, Korea}

\author{Tam, Pak-Hin Thomas}
\affiliation{School of Physics and Astronomy, Sun Yat-sen University, Guangzhou 510275, People’s Republic of China}

\author{Lili Yang}
\affiliation{School of Physics and Astronomy, Sun Yat-sen University, Guangzhou 510275, People’s Republic of China}

\author{Yudong Cui}
\affiliation{School of Physics and Astronomy, Sun Yat-sen University, Guangzhou 510275, People’s Republic of China}

\author[0000-0001-8922-8391]{Partha Sarathi Pal }
\affiliation{Laboratory for Space Research,  The University of Hong Kong, HK}

\author{Zhongli Zhang}
\affiliation{Shanghai Astronomical Observatory, Key Laboratory of Radio Astronomy, Chinese Academy of Sciences, Shanghai 200030, China}

\author{Bong Won Sohn}
\affiliation{Korea Astronomy and Space Science Institute, 776 Daedeok-daero, Yuseong-gu, Daejeon 34055, Korea}
\affiliation{Department of Astronomy and Space Science, University of Science and Technology, 217 Gajeong-ro, Daejeon, Korea}

\author{Koichiro Sugiyama}
\affiliation{National Astronomical Research Institute of Thailand (Public Organization), 260 Moo 4, T. Donkaew, A. Maerim, Chiang Mai, 50180, Thailand}

\author{Wen Chen}
\affiliation{Yunnan Observatories, Chinese Academy of Sciences, 650216 Kunming, Yunnan, China}
\affiliation{University of Chinese Academy of Sciences, 19A Yuquanlu, Beijing 100049, China}
\author{Longfei Hao}
\affiliation{Yunnan Observatories, Chinese Academy of Sciences, 650216 Kunming, Yunnan, China}



\begin{abstract}

Fermi J1544-0649 is a transient GeV source first detected during its GeV flares in 2017. Multi-wavelength observations during the flaring time demonstrate variability and spectral energy distribution(SED) that are typical of a blazar. Other than the flare time, Fermi~J1544-0649 is quiet in GeV band and looks rather like a quiet galaxy (2MASX J15441967-0649156) for a decade. Together with the broad absorption lines like feature we further explore "misaligned blazar scenario". We analysed the Very Long Baseline Array (VLBA) and East Asian VLBI Network (EAVN) data from 2018 to 2020 and discovered the four jet components from Fermi~J1544-0649. We found a viewing angle around 3.7$\degree$ to 7.4$\degree$. The lower limit of the viewing angle indicates a blazar with an extreme low duty cycle of gamma-ray emission, the upper limit of it support the "misaligned blazar scenario".
Follow-up multi-wavelength observations after 2018 show Fermi J1544-0649 remains quiet in GeV, X-ray and optical bands. Multi-messenger search of neutrinos is also performed, and an excess of 3.1 $\sigma$ significance is found for this source.

Key words: BL Lacertae objects:general, radio:general

\end{abstract}

\keywords{}


\section{Introduction} \label{sec:intro}

The current understanding of phenomenon of Active Galactic Nuclei (AGN) suggests that there is a super-massive black holes (SMBH) located at the center of galaxies, which can generate luminous emission over the whole electromagnetic spectrum. In some cases, an AGN could generate a relativistic jet. The unification scenario explains that when the jet direction nearly co-aligns with our line of sight, this AGN is called blazar with large variability at all wavelengths and usually accompanied by gamma-ray emission\citep{urry1995unified}. The largest identified source population in the $\gamma$-ray sky are blazars\citep{hartman1999third,abdo2009bright}, taking up to 53.4\% in the Fermi Large Area Telescope (Fermi-LAT) fourth source catalog of gamma-ray sources. 
Among these blazars, they are further grouped into BL Lacertae (BL Lac) objects and Flat Spectrum Radio Quasars (FSRQs), based on their optical properties. Also most of these gamma-ray observed BL Lacs have radio counterparts detected. Therefore, to understand their physical mechanisms one has to dig into their radio observations. 
By comparison, misaligned AGNs (MAGNs), with a jet pointed at larger angles to the viewer -- which are detected in large numbers at radio and optical frequencies -- are not so commonly observed in the $\gamma$-ray energy regime.

2MASX J15441967–0649156 is a galaxy with a low red shift of 0.171, and a GeV flare was found in May 2017 by FERMI-LAT gamma-ray telescope at this galaxy, not associated with any previously known gamma-ray source. During the flaring time, the X/$\gamma$-ray SED resemble those of extreme high-frequency-peaked BL Lac objects (EHBLs) and a fast X variation($<$ 1 hour) was also found (\citealt{bruni2018fermi}, \citealt{tam2020multi}).
However, unlike other known blazars, this galaxy remains quiet in X/$\gamma$-ray for a decade, is it a misaligned blazar? Strong optical variations was also found, which showed no connection with the X/$\gamma$-ray flare. Interestingly, Tam 2020 found two broad absorption lines (BALs) like feature during an optical flaring time, such BALs are usually found in quasar. If it is true, it also supports the "misaligned blazar scenario". To further explore this scenario, we try to observe its jet with the Very Long Baseline Array (VLBA) data and the East-Asian VLBI Network (EAVN\footnote{\url{https://radio.kasi.re.kr/eavn/main_eavn.php}}; for more information, see \citealt{10.1093/pasj/psu015,2016PASJ...68...72S,2018NatAs...2..118A}) data.

Such transient blazar with a large viewing angle may be a high-energy neutrinos source. The BAL like feature indicate clouds along our line of sight. According to the minijet/wobbling model, the agn is capable to throw out some material with a large inclination angle to the earth. Once the collision between jet and the clouds happen, it would be an ideal place for neutrino production via photonmeson interaction. Therefore we searched ten years' archival neutrino data and found 4 neutrino events with a 3.1 $\sigma$ significance excess.

The paper is organized as follows. We give radio observations and data reduction in section 2. A result of radio data analysis is described in section 3. We provide an account of the multi-wavelength and multi-messenger observations in section 4. The discussion and conclusions are given in section 5 and 6.

\section{Radio Observations and Data Reduction}

\subsection{VLBA Observations at 5 and 8 GHz}\label{VLBA}

The VLBA observations (project code: BT146, PI: P.H.T. Tam) were done on February 10 and May 20 2019 at 4.87 GHz, and on February 11 and May 21 2019 at 8.37 GHz.
The duration of each segment was about 1.5 hr.
Due to the weakness of the source, we used phase referencing technique with a 3.5-min cycle of ``calibrator (1min)-target (2.5min)" at 5 GHz and a 1.5-min cycle of ``calibrator (0.5min)-target (1min)" at 8 GHz. 
The phase calibrator is chosen to be J1543-0757 which is 1.18$\degree$ from Fermi J1544-0649.
The data were recorded at 2048 Mbps rate, with 2 polarizations, 2 intermediate frequency channels (IFs) per polarization, and 128 MHz bandwidth per IF.
The experiment set-ups are summarized in Table \ref{tab:obs}.
The data were correlated using the DiFX software correlator \citep{2007PASP..119..318D,2011PASP..123..275D} at Socorro with an averaging time of 2s, 256 frequency channels per IF and uniform weighting.

We calibrated the data in the US National Radio Astronomy Observatory (NRAO) Astronomical Imaging Processing System (AIPS; \citealt{2003ASSL..285..109G}).
A priory amplitude calibration of the visibility is carried out using the system temperatures and antenna gains measured at each station during the observations.
The dispersive delays caused by the ionosphere were corrected according to a map of total electron content provided by Global Positioning System (GPS) satellite observations.
Phase errors due to the antenna parallactic angle variations were removed.
The instrumental single-band delays and phase offsets were corrected using 2-min observational data of the calibrator 3C 345.
After inspecting the data and flagging, global fringe-fitting was performed on the phase-referencing calibrator (J1543-0757) with a 0.5-min solution interval and a point-source model \citep{1995ASPC...82..189C} by averaging over all the IFs.
We then applied the phase corrections from the calibrator source to interpolate to the target source.

\begin{deluxetable*}{cccccc}
\centering
\tablenum{1}
\tablecaption{Details of the VLBA and KaVA observations\label{tab:obs}}
\tabletypesize{\small}
\tablewidth{0pt}
\tablehead{
\colhead{Date} & \colhead{$\rm \nu_{obs}$} & \colhead{Time} & \colhead{Rate}  & \colhead{Project code} & \colhead{Participating stations} \\
\colhead{(yyyy-mm-dd)} & \colhead{(GHz)} & \colhead{(min)} & \colhead{(Mbps)} & \colhead{} & \colhead{}
}
\decimalcolnumbers
\startdata
2018-10-16 & 4.34 & 10 & 1024 & SB072B0 & VLBA (except KP)  \\
2018-10-16 & 7.62 & 10 & 1024 & SB072B0 & VLBA (except KP) \\
2019-02-10 & 4.87 & 75 & 2048 & BT146C1 & VLBA (except MK) \\
2019-02-11 & 8.37 & 77 & 2048 & BT146X1 & VLBA (except MK, OV, BR) \\
2019-05-20 & 4.87 & 75 & 2048 & BT146C2 & VLBA (except SC) \\
2019-05-21 & 8.37 & 75 & 2048 & BT146X2 & VLBA (except PT, SC) \\
2020-09-14 & 6.73 & 40 & 1024 & A20T3C  & MIZ, OGA, ISG, KUS, TAK, HIT, YAM, SHA, KUN \\
\enddata
\tablecomments{The columns give the following: (1) observing date; (2) observation frequency; (3) on source time for Femi J1544$-$0649; (4) data rate; (5) project code; (6) participating stations: The VLBA stations that were not used in individual observations are shown in brackets: BR, Brewster; PT, Pie Town; KP, Kitt Peak; OV, Owens Valley; MK, Mauna Kea; SC, St Croix. The abbreviations for the EAVN stations (see Section \ref{EAVN} for the explanation on the telescope codes).}
\end{deluxetable*}

We firstly imaged the calibrator J1543-0757.
The calibrator shows two extended components separated by
50 mas, with the northern component $\sim$3 three times brighter.
We iteratively ran model fitting with point sources and self-calibration in Difmap \citep{1994BAAS...26..987S}, fringe fitting, and self-calibration to remove its structure-dependent phase errors in AIPS.
We also ran amplitude self-calibrations on the calibrator data and transferred the solutions to the target data.
The visibility data were fitted with three circular Gaussian components in Difmap using the MODELFIT program to minimize the potential deconvolution errors of CLEAN.
The typical uncertainties of total flux density are less than 10\% and are mainly contributed by the visibility amplitude calibration errors and the antenna gain calibration errors. 
The uncertainty in the fitted component size is less than 15\% of the
deconvolved size of the fitted Gaussian model. We estimated the errors in the best-fitting positions of the Gaussian jet components, 20\% of the component size convolved with the synthesized beam size \citep{2021MNRAS.506.1609C}.

\subsection{Archival VLBA observation}

To further study this source, we obtained the fully calibrated VLBI data taken in astrogeo database\footnote{VLBA calibrator survey data base is maintained by Leonid Petrov, http://astrogeo.org/.}.
The source is observed simultaneously at dual frequencies (5 and 8 GHz) with two 5-min scans in snapshot mode on 2018 October 16, which is usually used for astrometric and geodetic studies.
The project code is SB072B0.
The data was imaged and modeled in Difmap to quantitatively describe the emission structure.
To reduce errors due to differences in data quality (in practice, the difference in data quality is not much), we used the same model fitting uncertainties for all VLBA data.
Further information regarding the observation can be found in Table \ref{tab:obs}.
\\ \hspace*{\fill} \\

\subsection{EAVN observation at 6.7 GHz}\label{EAVN}

The observation was performed with the EAVN on 2020 September 14.
The project code is a20t3c, one epoch of the EAVN C-band test observations.
The observation also used phase referencing mode with a 4-min cycle of "calibrator (1 min), target (2 min), and antenna slewings of 30 sec each" at 6.7 GHz.
The experiment consisted of nine antennas from the following: Yamaguchi 32 m (YAM), Hitachi 32 m (HIT), Shanghai 25 m (SHA), Kunming 40m (KUN),  Takahagi 32m (TAK), KVN Ulsan 21m (KYS) and VERA (three 20 m stations: Mizusawa, Ogasawara, and Ishigaki).
The data was recorded in 16 IFs with a bandwidth of 16 MHz per IF in left circular polarization, resulting in a total bandwidth of 256 MHz (6,600-6,856 MHz) and a sampling rate of 1024 Mbps and had an on-source time for Femi J1544-0649 of 40 min.
The experiment set-ups is summarized in Table \ref{tab:obs}.
NRAO 512 and NRAO 530 were observed as calibrators.
The correlation was carried out in the Korea–Japan Correlation
Center (KJCC) at Daejeon, Korea \citep{2014AJ....147...77L}, with an integration time for each visibility output of 1.63 s. 

\begin{figure*}
\centering
\includegraphics[width=0.45\textwidth]{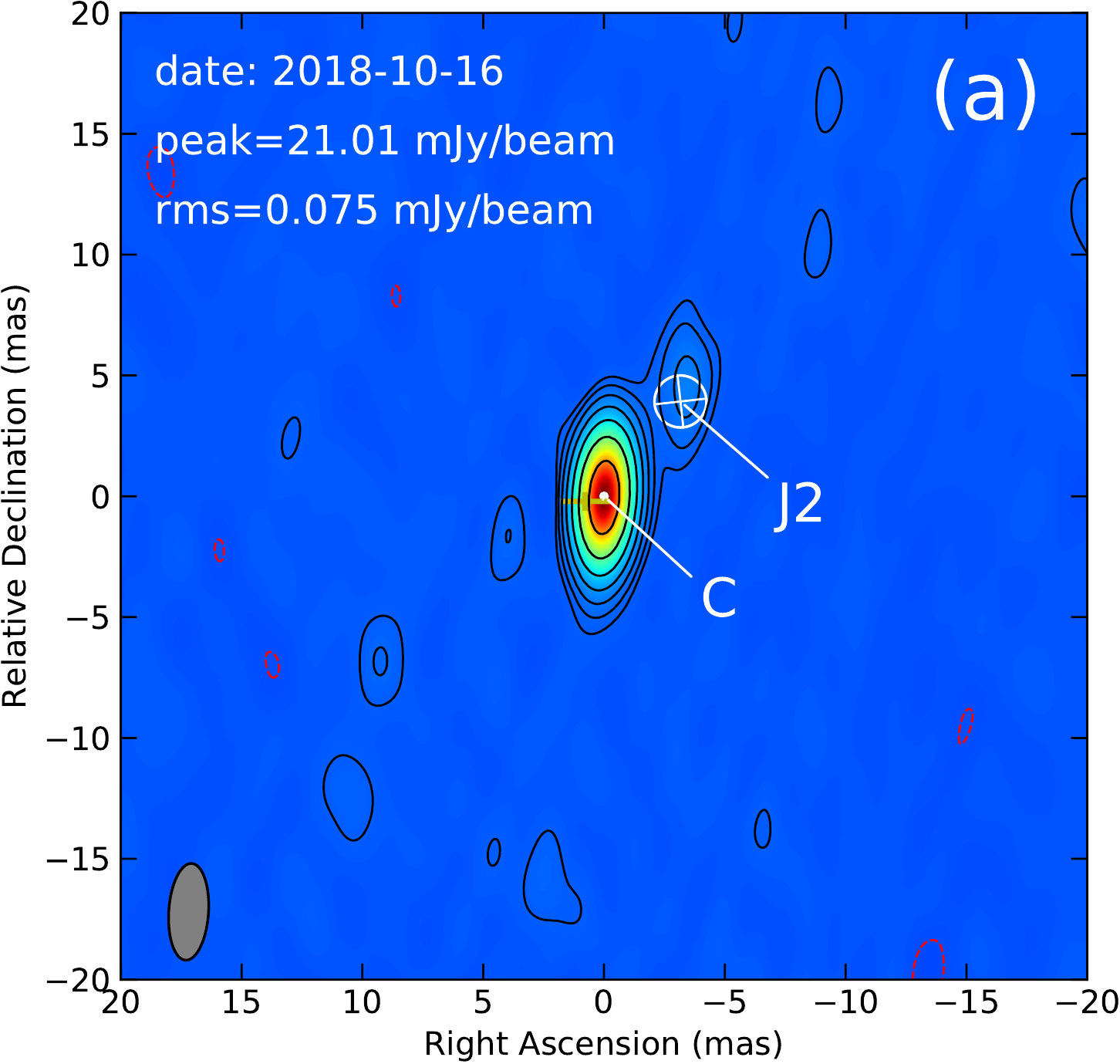}
\includegraphics[width=0.45\textwidth]{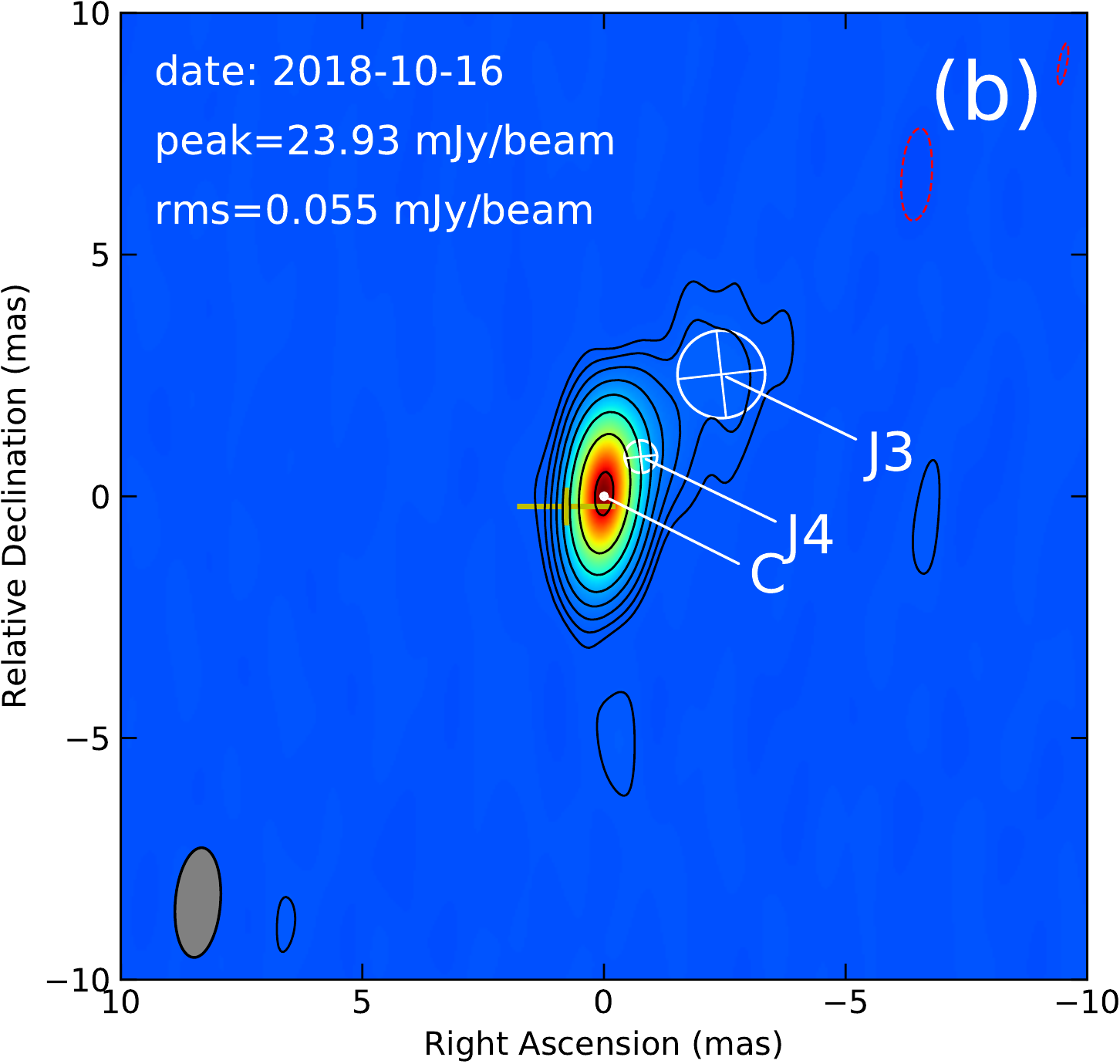}\\
\includegraphics[width=0.45\textwidth]{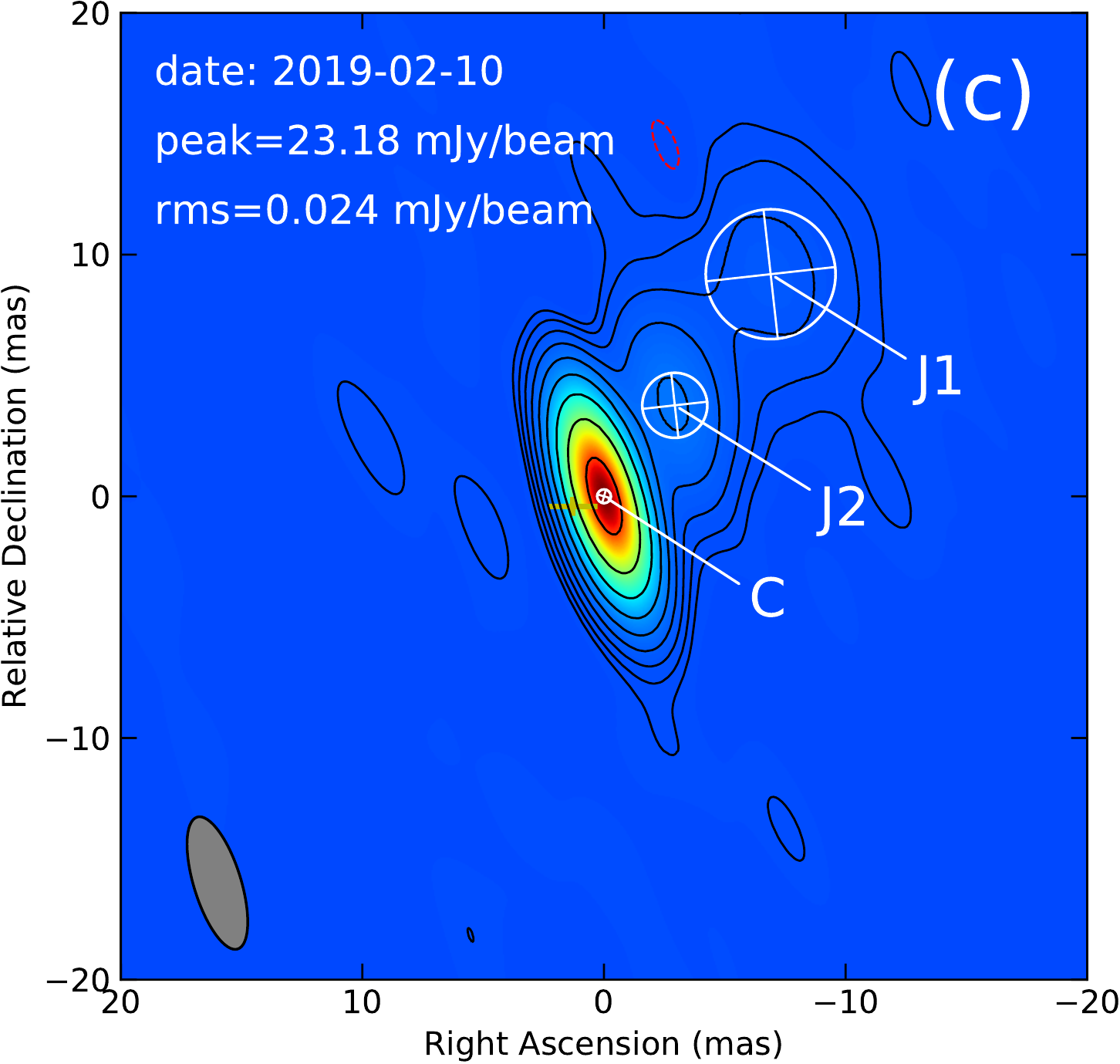}
\includegraphics[width=0.45\textwidth]{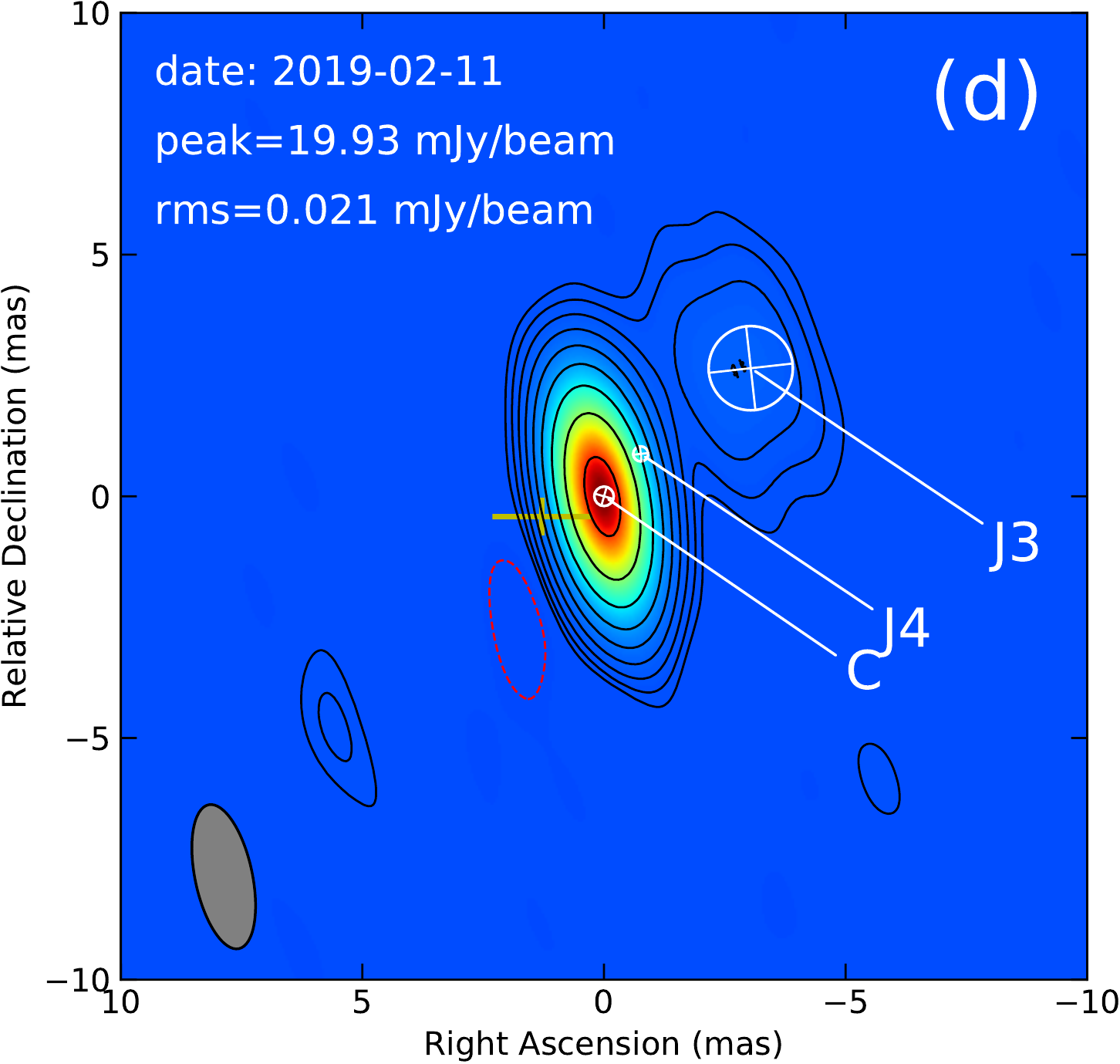}\\
\includegraphics[width=0.45\textwidth]{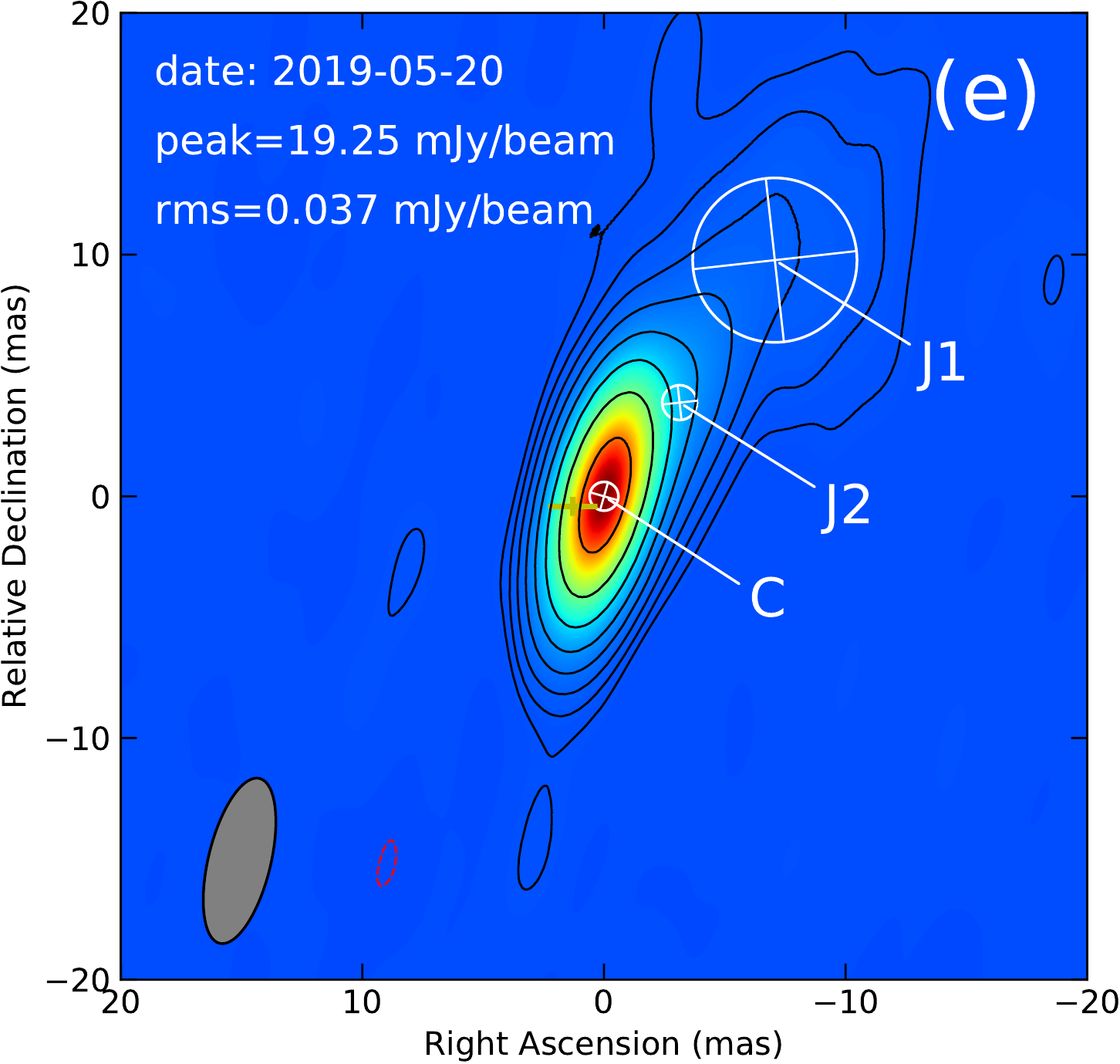}
\includegraphics[width=0.45\textwidth]{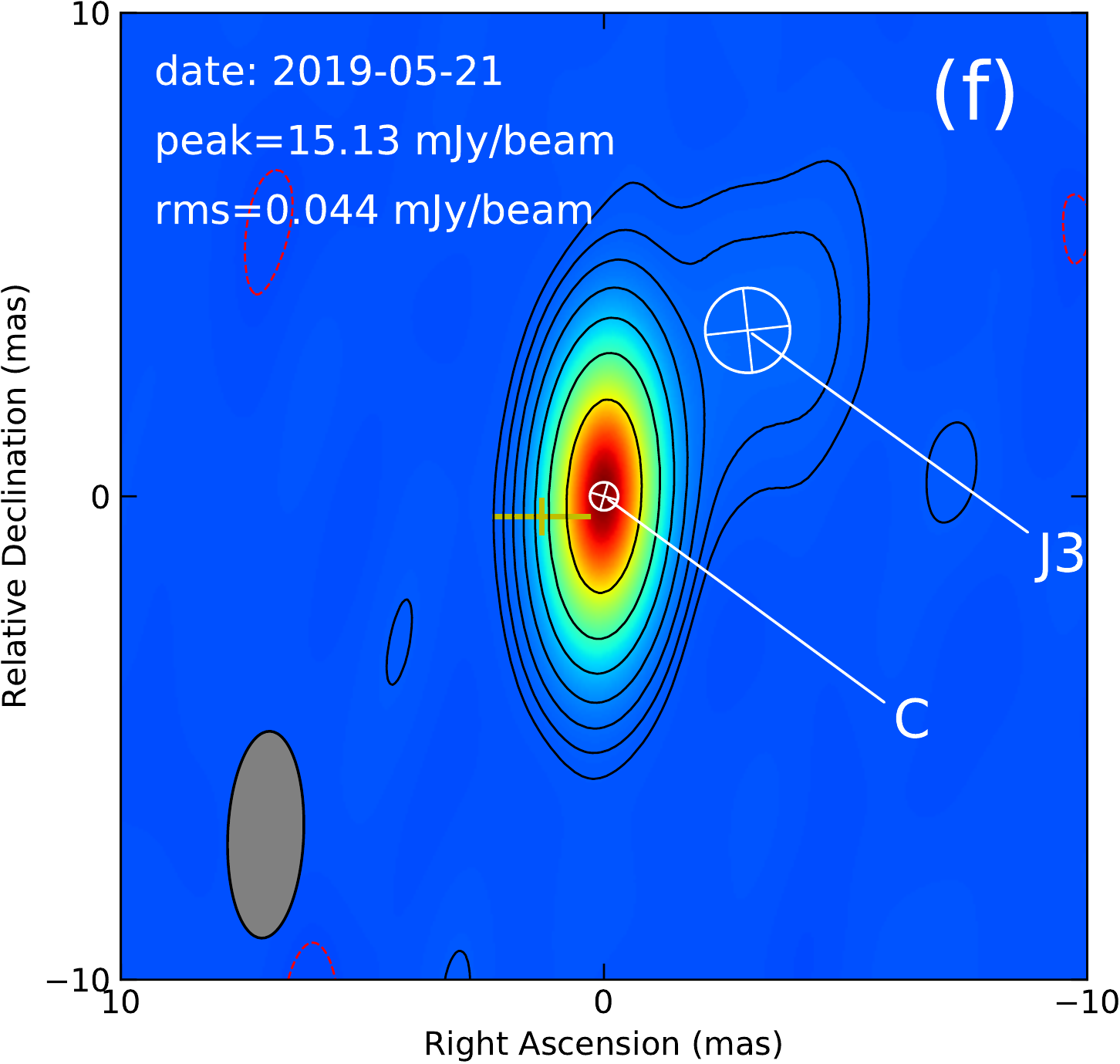}\\
\caption{\label{fig:image-VLBA}Naturally weighted images at 5 GHz (left) and 8 GHz (right) from the VLBA observations. The restoring beam is shown in the lower-left side of each panel. The contour levels start at three times the rms noise value, and positive levels increase by a factor of 2. The yellow cross marks the position derived from the Gaia astrometry catalog, corresponding to the location of the AGN. The image parameters are listed in Table \ref{tab:img}.}
\end{figure*}

The correlated data was analyzed with AIPS and Difmap, following the standard phase-referencing VLBI data reduction procedures (the same procedure described in Section \ref{VLBA}).
The instrumental single-band delays and phase offsets were corrected using 2-min observational data of the calibrator NRAO 530.
The bandpass character was calibrated with the NRAO530 as well.
Scaling flux and the gain curve calibration were completed by a priori method for KYS, VERA, HIT, SHA, and KUN. 
On the other hand, YAM and TAK were calibrated by a self-calibration way by applying to a point-like source NRAO512.
Note that for KYS, a special amplitude correction factor of 1.3 is needed, as suggested by \citet{2015JKAS...48..229L}.
An amplitude calibration uncertainty of 15\% is estimated for the EAVN data \citep{2014PASJ...66..103N}. 
Most stations had good fringe detection for calibrator J1543-0757 in this experiment.
After a global fringe search, the data was averaged in 30 s intervals and across the entire bandwidth for Gaussian model fitting.

\section{Radio Interference Result}

\subsection{Parsec-scale radio morphology}

\begin{figure}
    \centering
    \includegraphics[width=0.4\textwidth]{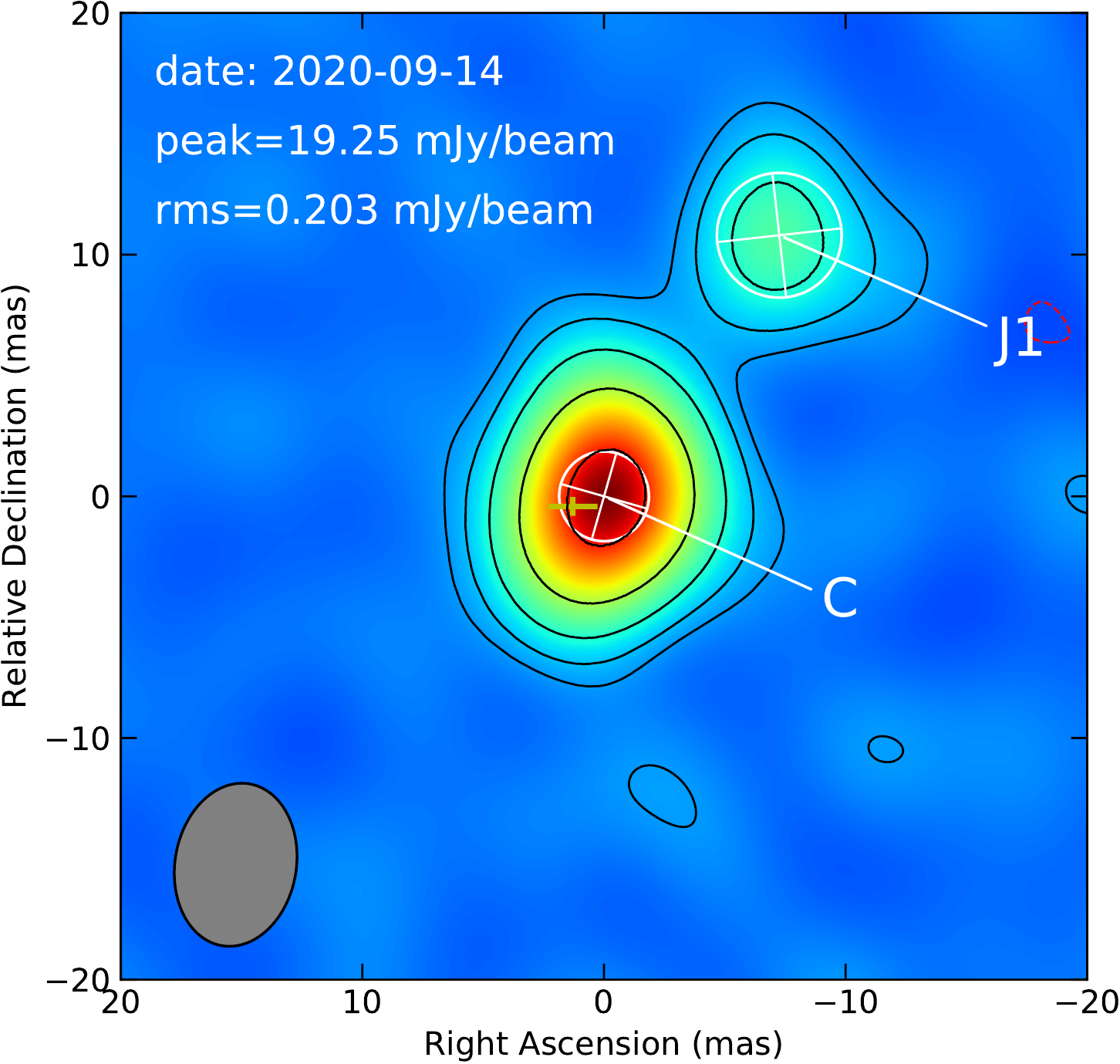}
    \caption{A 6.7 GHz EAVN image of Fermi J1544$-$0649 is made with naturally weighting. The restoring beam is shown in the lower-left side of each panel. The contour levels start at three times the rms noise value (0.203 mJy beam$^{-1}$), and positive levels increase by a factor of 2. The yellow-colored cross marks the position derived from the Gaia astrometry catalog, corresponding to the location of the AGN.  The image parameters are listed in Table 2.}
    \label{fig:image-KaVA}
\end{figure}

Figures \ref{fig:image-VLBA} and \ref{fig:image-KaVA} show the naturally weighted total intensity images obtained from the VLBA and KaVA data.
The elliptical Gaussian restoring beam is indicated in the bottom-left corner of each image in Fig. \ref{fig:image-VLBA} and \ref{fig:image-KaVA}.
The parameters of all the individual images and contour levels for all observations are listed in Table \ref{tab:img} and results from Gaussian model fitting of clearly discernible components C and J1 - J4 are presented in Table \ref{tab:model}.
The source shows a compact core–jet structure with a compact component appearing at one end (identified as the core, $\alpha$\footnote{Spectral index $\alpha$ is defined as $\rm S_{\nu}$ $\propto$ $\rm \nu^{\alpha}$ .} = 0.10) and the jet component is labeled as J.
The location and the full width at half maximum (FWHM) size of the components are also indicated in the images. 
VLBI phase-referencing allows us to measure the extract absolute positional information for the target source: RA = 15$\rm ^{h}$44$\rm ^{m}$19$\rm ^{s}$.65305, Dec. = $-$06$\degree$49$\rm ^{'}$15$\rm ^{''}$.3957 (J2000, $\rm \sigma_{RA}$ = 0.53 mas, $\rm \sigma_{Dec}$ = 1.20 mas).
The optical centroid, reported by the Gaia DR2 \citep{2018A&A...616A...1G}, is marked as a yellow cross (J2000, RA = 15$\rm ^{h}$44$\rm ^{m}$19$\rm ^{s}$.65305, Dec. = $-$06$\degree$49$\rm ^{'}$15$\rm ^{''}$.3988, $\rm \sigma_{RA}$ = 1.02 mas, $\rm \sigma_{Dec}$ = 0.39 mas).
With respect to the optical centroid, the radio peak has an offset of about 1 mas, consistent with our VLBI position.

The images (left panel in Fig. \ref{fig:image-VLBA}) at 5 GHz show a well-resolved core–jet structure extending to the north-west (NW) direction with two jet components located about 4.5 mas (J2) and 12 mas (J1) of
the core (C), at a position angle of about $-$40$\degree$ (counterclockwise from north).
The finest angular resolution was provided by the VLBA observations at 8 GHz.
Two new jet components (J3 and J4) were detected in the inner region ($<$ 5 mas).
J4 component was not detected in the last epoch due to St. Croix being in maintenance and lack of sufficient angular  resolution.
The four components (J1 - J4) are aligned in the NW direction, at a position angle of $\sim$ $-$40$\degree$.
Owing to the limited $uv$ coverage, only the core and the outer jet component J1 were detected in the 6.7 GHz EAVN image (Fig. \ref{fig:image-KaVA}).

\begin{figure}[!htp]
    \flushright
    \includegraphics[width=0.45\textwidth]{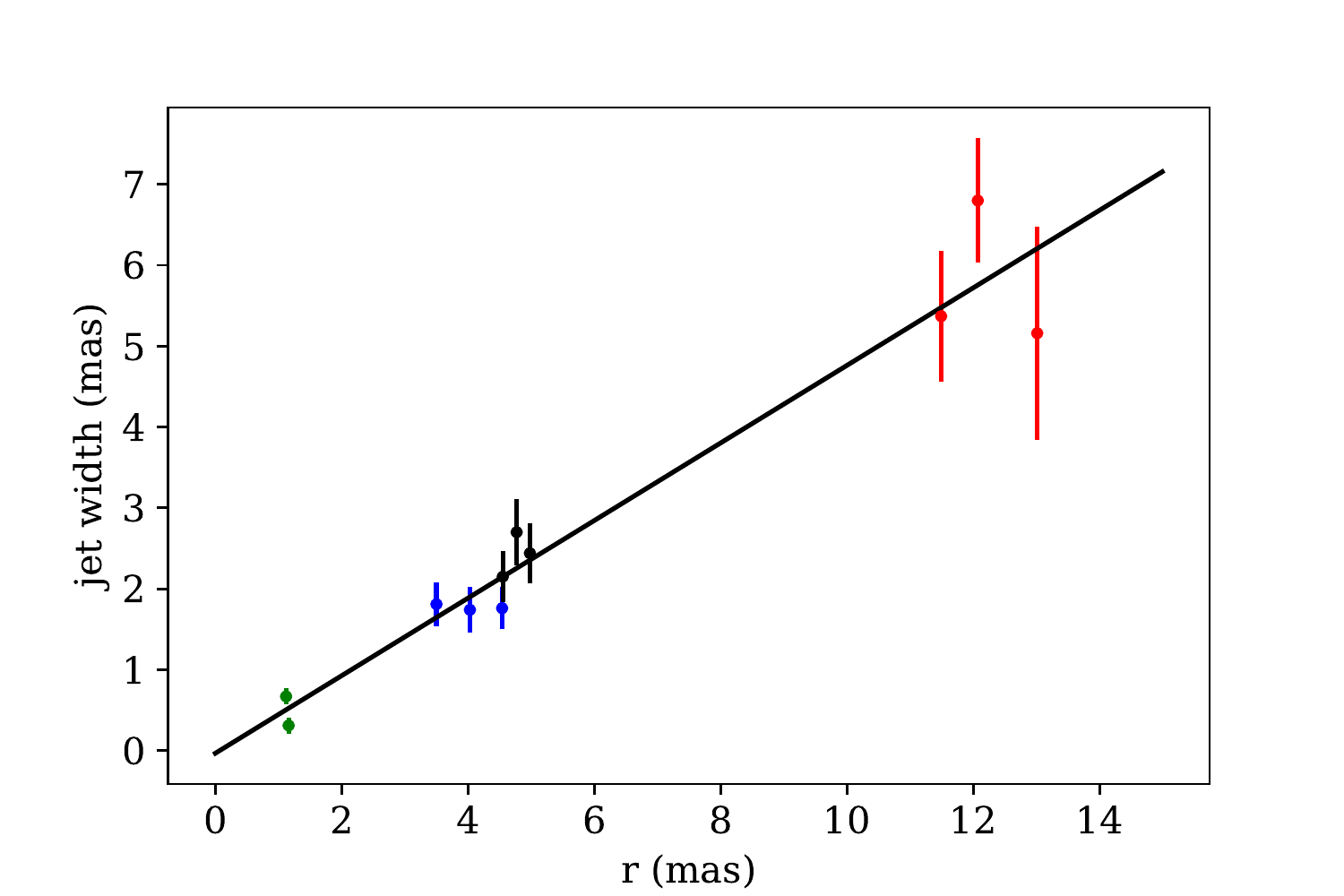}
    \caption{Opening angle of the jet body.  The data points are derived from our observations and presented in Table \ref{tab:model}. From bottom left to top right are J4, J3, J2, J1.}
    \label{fig:jet-width}
\end{figure}

The sizes of all jet components increase with the radial distance from the core. 
The outermost jet component J1 is the most extended with the largest size, and the innermost J4 has the smallest size. 
Therefore a conical geometry could be a good description of the Fermi J1544$-$0649 jet. 
Since the jet body is well aligned, the opening angle of the jet can be represented by all jet components.
Figure \ref{fig:jet-width} shows the jet width (D) versus radial distance of the jet component from the core (r) for all the data.
The jet components size from the model fitting of the visibility data with circular Gaussian models can be used as the jet width and are listed in Table \ref{tab:model}.
Owing to the errors arising from the model fitting of their corresponding brightness distributions, the data points of each component show scattered along the jet width axis.
The jet width is well fitted with a linear function, supporting the inference of a conical jet body.
The projected opening angle of the jet beam $\rm \alpha_{pro}$ = 2arctan(D/2r) = 26.9$\degree$ $\pm$ 2.2$\degree$, similar to the mean value of 22.4$\degree$ for BL Lacs in the MOJAVE sample \citep{2017MNRAS.468.4992P}.

\subsection{Jet parameters}

We estimated the brightness temperature of the core components using the results of our brightness distribution model fitting (e.g., \citealt{1996ApJ...461..600G,2005AJ....130.2473K})
\begin{equation}
  T_{\rm b} = 1.22\times10^{\rm 12}\frac{S}{ \nu^{2}d^{2}}(1+z),
\end{equation}
where $S$ is the flux density of the core in Jy, $z$ is the redshift, $\nu$ is the observing frequency in GHz, and $d$ is the fitted Gaussian size (FWHM) of the component in mas. 
We use the highest core brightness temperature of 3.5 $\times$ 10$^{10}$ K, comparable to the equipartition value \citep{1995ASPC...82..189C}, in the following calculation, since epoch 2018 October 16 at 8 GHz gives the best resolution.
A study of a large sample of radio-loud AGNs shows that core brightness temperature is about 10$^{10}$ -- 10$^{13}$ K \citep{2020ApJS..247...57C}, indicating the T$_{b}$ of Fermi J1544$-$0649 at the low-end tail of the brightness temperature distribution.

\begin{deluxetable*}{cccccc}
\centering
\tablenum{2}
\tablecaption{Image parameters in Fig. \ref{fig:image-VLBA} and \ref{fig:image-KaVA}.\label{tab:img}}
\tabletypesize{\small}
\tablewidth{0pt}
\tablehead{
\colhead{Figure label} & \colhead{Date} & \colhead{Band} & \colhead{S$_{\rm peak}$}  & \colhead{Contours} & \colhead{Beam FWHM and PA} \\
\colhead{} & \colhead{(yyyy-mm-dd)} & \colhead{(GHz)} & \colhead{(mJy beam$^{-1}$)} & \colhead{mJy beam$^{-1}$} & \colhead{(mas$\times$mas, $\degree$)}
}
\decimalcolnumbers
\startdata
Fig. \ref{fig:image-VLBA}(a) & 2018-10-16 & 4.34 & 21.01 & 0.22$\times$(-1,1,2,...,64) & 4.03$\times$1.55, $-$2.78$\degree$  \\
Fig. \ref{fig:image-VLBA}(b) & 2018-10-16 & 7.62 & 23.93 & 0.17$\times$(-1,1,2,...,128) & 2.28$\times$0.94, $-$4.69$\degree$  \\
Fig. \ref{fig:image-VLBA}(c) & 2019-02-10 & 4.87 & 23.18 & 0.08$\times$(-1,1,2,...,128) & 5.71$\times$1.95, 16.92$\degree$  \\
Fig. \ref{fig:image-VLBA}(d) & 2019-02-11 & 8.37 & 19.93 & 0.09$\times$(-1,1,2,...,128) & 3.06$\times$1.19, 11.93$\degree$  \\
Fig. \ref{fig:image-VLBA}(e) & 2019-05-20 & 4.87 & 19.25 & 0.11$\times$(-1,1,2,...,128) & 6.99$\times$1.79, $-$16.77$\degree$  \\
Fig. \ref{fig:image-VLBA}(f) & 2019-05-21 & 8.37 & 15.13 & 0.13$\times$(-1,1,2,...,64) & 4.28$\times$1.57, $-$2.42$\degree$  \\
Fig. \ref{fig:image-KaVA}    & 2020-09-14 & 6.73 & 19.25 & 0.61$\times$(-1,1,2,...,16) & 6.79$\times$5.02, $-$9.78$\degree$  \\
\enddata
\tablecomments{The columns give the following: (1) figure label; (2) observation date; (3) observation frequency; (4) peak specific intensity; (5) contours level, the lowest contour level corresponding to the 3 times of off-source rms noise in the clean image ; (6) major axis and minor axis of the restoring beam, and the position angle of the major axis, measured from north to east.}
\end{deluxetable*}

\begin{deluxetable*}{cccccccc}
\centering
\tablenum{3}
\tablecaption{Gaussian model-fitting results of the components in Femi J1544$-$0649.\label{tab:model}}
\tabletypesize{\small}
\tablewidth{0pt}
\tablehead{
\colhead{Date} & \colhead{$\rm \nu_{obs}$} & \colhead{Name} & \colhead{S$_{\rm peak}$}  & \colhead{S$_{\rm tot}$} & \colhead{R} & \colhead{P.A.} & \colhead{$\rm \Theta$} \\
\colhead{(yyyy-mm-dd)} & \colhead{(GHz)} & \colhead{} & \colhead{(\textbf{mJy beam$^{-1}$})} & \colhead{mJy} & \colhead{(mas)} & \colhead{$\degree$)} & \colhead{(mas)}
}
\decimalcolnumbers
\startdata
2018-10-16 & 4.34 & C & 21.01 & 22.01$\pm$2.20 & ...            & ...      & 3.71$\pm$0.56 \\
          &      & J2& 0.94  & 1.12$\pm$0.11  & 4.55$\pm$0.43  & $-$38.82 & 2.15$\pm$0.32 \\
2018-10-16 & 7.62 & C & 23.93 & 23.22$\pm$2.32 & ...            & ...      & 0.14$\pm$0.02 \\
          &      & J4& 7.67  & 1.98$\pm$0.20  & 1.12$\pm$0.13  & $-$42.25 & 0.67$\pm$0.10 \\
          &      & J3& 0.46  & 1.31$\pm$0.13  & 3.50$\pm$0.36  & $-$43.33 & 1.81$\pm$0.27 \\
2019-02-10 & 4.87 & C & 23.18 & 24.19$\pm$2.49 & ...            & ...      & 0.58$\pm$0.56 \\
          &      & J2& 1.32  & 2.32$\pm$0.23  & 4.77$\pm$0.54  & $-$37.04 & 2.70$\pm$0.41 \\ 
          &      & J1& 0.41  & 1.52$\pm$0.15  & 11.49$\pm$1.07 & $-$36.67 & 5.37$\pm$0.81 \\
2019-02-11 & 8.37 & C & 19.93 & 20.91$\pm$2.09 & ...            & ...      & 0.40$\pm$0.06 \\
          &      & J4& 6.23  & 0.54$\pm$0.05  & 1.16$\pm$0.06  & $-$40.22 & 0.31$\pm$0.05 \\
          &      & J3& 0.51  & 0.91$\pm$0.09  & 4.03$\pm$0.35  & $-$48.66 & 1.74$\pm$0.28 \\
2019-05-20 & 4.87 & C & 19.25 & 19.87$\pm$1.99 & ...            & ...      & 1.19$\pm$0.18 \\
          &      & J2& 4.50  & 1.74$\pm$0.17  & 4.98$\pm$0.49  & $-$38.93 & 2.44$\pm$0.37 \\
          &      & J1& 0.50  & 2.62$\pm$0.26  & 12.07$\pm$1.76 & $-$35.08 & 8.80$\pm$1.32 \\
2019-05-21 & 8.37 & C & 15.13 & 16.16$\pm$1.62 & ...            & ...      & 0.58$\pm$0.09 \\
          &      & J3& 0.58  & 2.07$\pm$0.21  & 4.53$\pm$0.35  & $-$40.34 & 1.76$\pm$0.26 \\
2020-09-14 & 6.73 & C & 11.62 & 16.67$\pm$2.50 & ...            & ...      & 3.71$\pm$0.56 \\
          &      & J1& 3.22  & 4.05$\pm$0.61  & 13.01$\pm$1.03 & $-$33.90 & 5.16$\pm$0.77 \\
\enddata
\tablecomments{The columns give the following: (1) observing date; (2) observation frequency; (3) component name; (4) peak specific density; (5) model flux density; (6) separation from the core; (7) position angle with respect to the core, measured from north through east; (8) size.}
\end{deluxetable*}

\begin{figure}[!htp]
    \centering
    \includegraphics[width=0.4\textwidth]{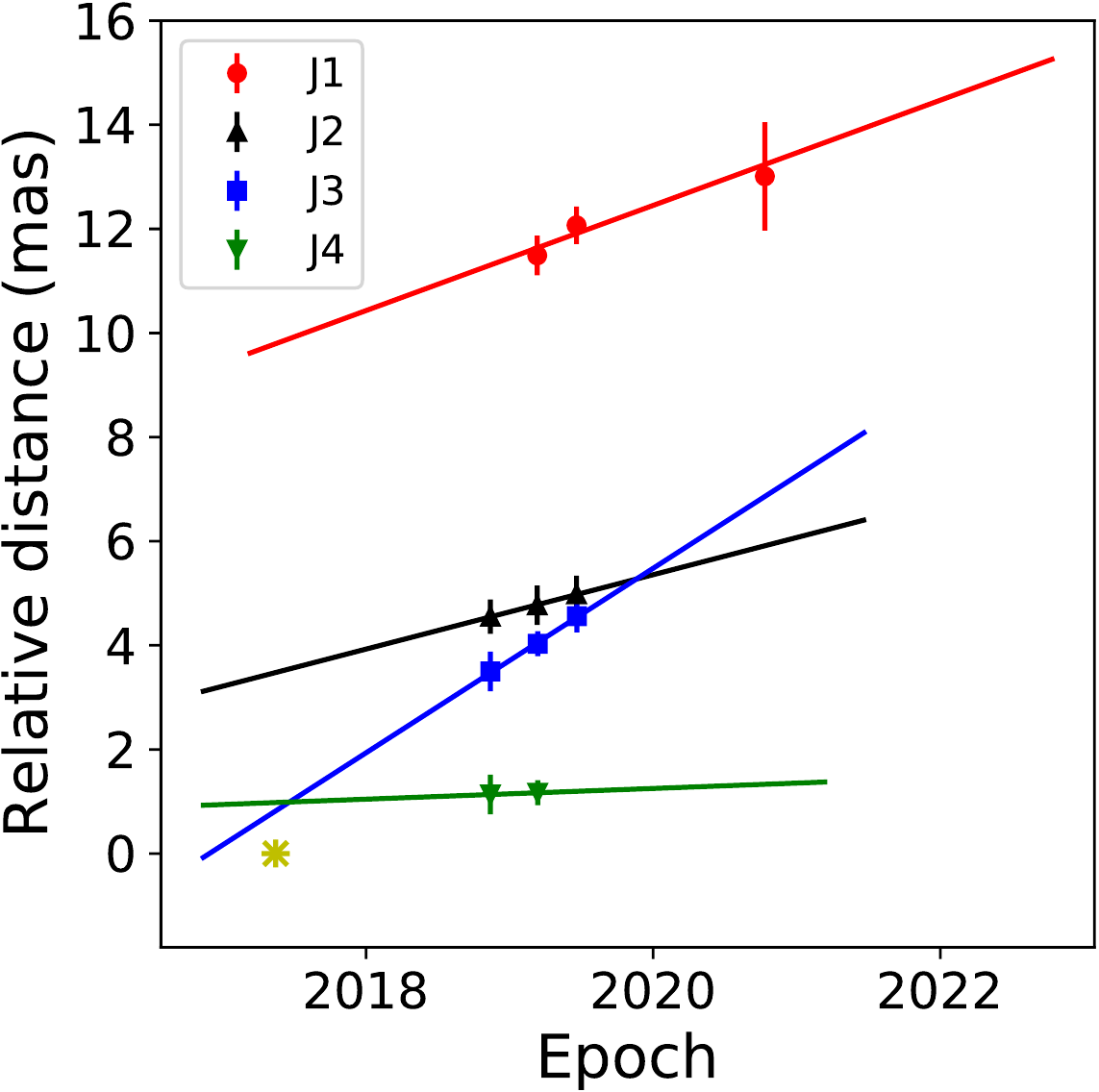}
    \caption{The proper motions of jet components are determined by the least-squares linear fit to the component positions as a function of time. The asterisk represents the time of the GeV flare.}
    \label{fig:pm}
\end{figure}

Figure \ref{fig:pm} shows the radial distance of the jet component from the radio core versus observing time.
All the jet components move along a constant radial direction.
In order to quantitatively study the jet kinematics, we used a non-acceleration, two-dimensional vector fit to the jet component’s position with time \citep{2016AJ....152...12L}.
The jet speeds $\beta_{\rm app}$ of J1, J2, J3, and J4 are 1.01$\pm$0.62 mas yr$\rm ^{-1}$ (9.50$\pm$5.84 c), 0.73$\pm$0.02 mas yr$\rm ^{-1}$ (6.88$\pm$0.19 c), 1.63$\pm$0.24 mas yr$\rm ^{-1}$ (15.33$\pm$2.21 c), and 0.06 mas yr$^{-1}$ (0.57c), respectively.
J4 only appears in the first two epochs of 8 GHz data; since it is very close to the core, the model fitting of J4 is much affected by the mixture with the core emission.
In the following discussion, two methods of calculation are given and we will use the fastest measured radial, non-accelerating apparent jet speed estimated for J3 as a representative of the jet speed.

\begin{figure*}[!htp]
    \centering
    \includegraphics[width=0.45\textwidth]{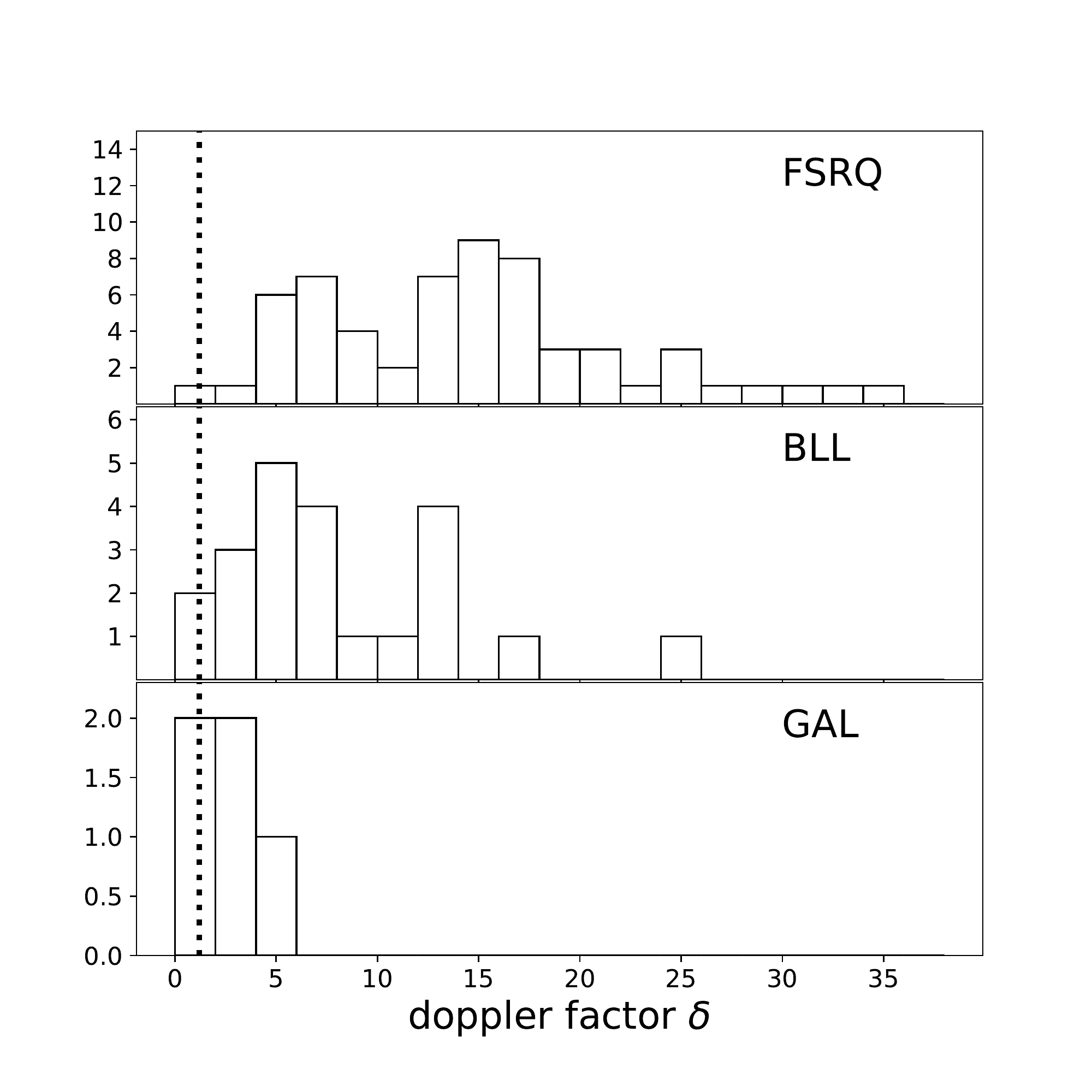}
    \includegraphics[width=0.45\textwidth]{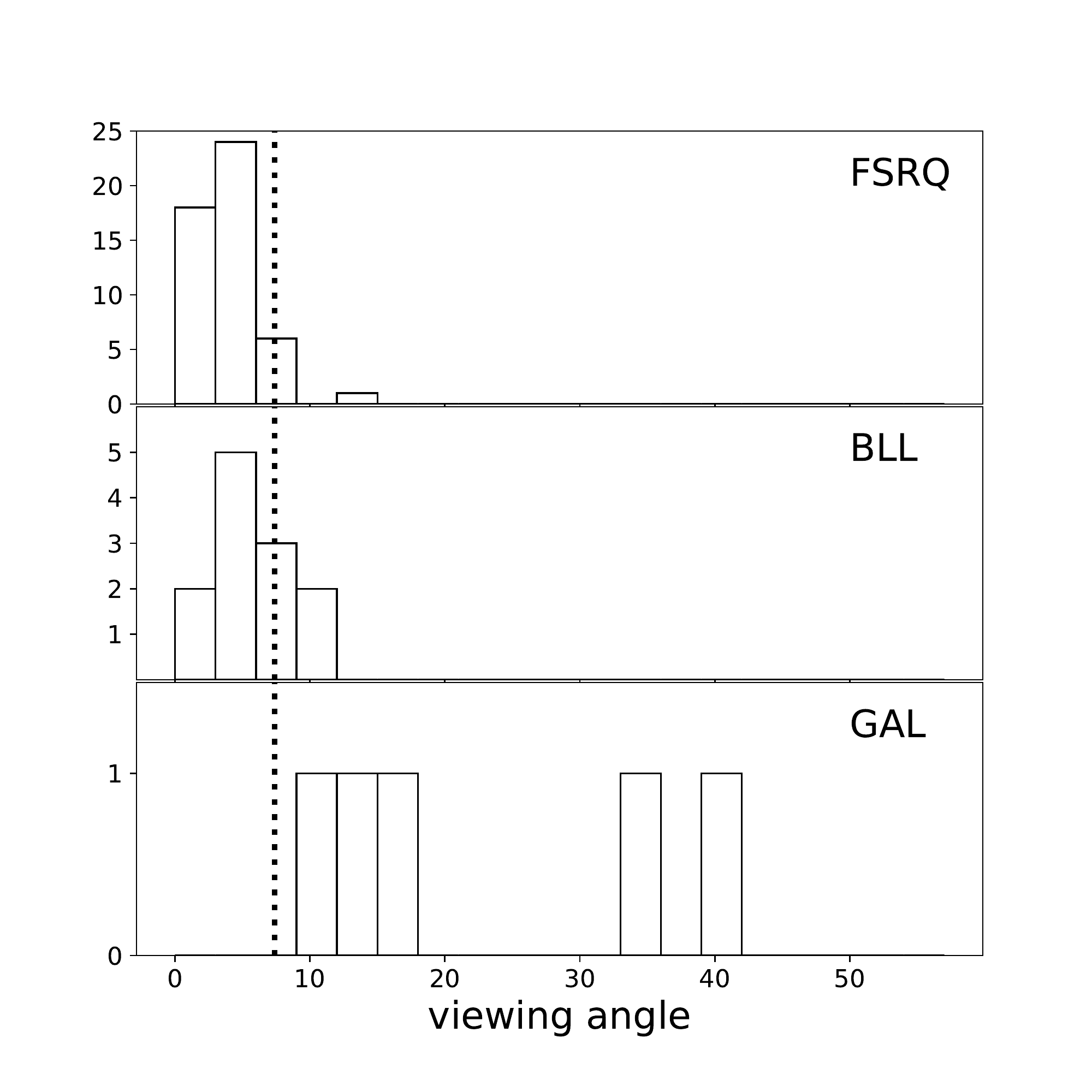}
    \caption{left panel: statistical distribution of doppler factor. right panel: statistical distribution of viewing angle. The dashed line on behalf of fermi 1544-0649. The statistical distributions are from Metsahovi Radio Observatory monitoring list. The observed doppler factor of Fermi 1544-0649 during its non-flare time is as low as those of a galaxy. The viewing angle is larger than most of the blazar.}
    \label{fig:flux_dopp}
\end{figure*}

In the first method, the intrinsic opening angle ($\rm \alpha_{int}$) and viewing angle ($\theta$) can be calculated using the following relations\textbf{\citep{hovatta2009doppler,2017MNRAS.468.4992P}}:
\begin{equation}
    \theta = \rm arctan\frac{2\beta_{app}}{\beta_{app}^{2}+\delta^{2}-1}
\end{equation}
\begin{equation}
    \alpha_{\rm int} = \rm 2arctan(tan(\alpha_{\rm pro}/2)sin\theta).
\end{equation}
For $\beta_{\rm app}$ and $\rm \alpha_{pro}$ we used the fastest measured radial non-accelerating apparent jet 15.33$\pm$2.21 and the projected opening angle 26.9$\degree$ $\pm$ 2.2$\degree$, respectively.Doppler boosting factor is conventionally associated with relativistic jets seen at a small viewing angle.
The jet Doppler factor can be inferred from a single epoch VLBI observation
\begin{equation}
  \delta = \frac{T_{\rm b,core}}{T_{\rm int}},
\end{equation}
where T$\rm _{b,core}$ = 3.5 $\times$ 10$^{10}$ K is the highest core brightness temperature and T$\rm _{int}$ is the intrinsic brightness temperature. Here we assume the T$\rm _{int}$ = 3 $\times$ 10$^{10}$ K proposed by \citet{2006ApJ...642L.115H}, which is close to the value expected for equipartition, when the energy in the radiating particles equals the energy stored in the magnetic fields. The Doppler factor $\delta$ is 1.2$\pm$0.2.
The intrinsic jet opening angle is $\rm \alpha_{int}$ = 3.55$\degree$ $\pm$ 0.6 $\degree$ and viewing angle $\theta$ = 7.42$\degree$ $\pm$ 1.68$\degree$, reflecting a high degree of jet collimation.
We combined the Doppler factor and viewing angle of Fermi 1544-0649 with the well-monitored AGN from the Metsahovi Radio Observatory monitoring list in Fig. \ref{fig:flux_dopp} \citep{hovatta2009doppler}. It shows that the observed Doppler factor of Fermi 1544-0649 during its non-flare time is as low as those of a galaxy and the viewing angle is larger than most of the blazar.

In the second method, it is possible that we underestimate the Doppler boosting because the intrinsic brightness temperature of this source may be lower than the normal blazars.
In this case, we can assume that the viewing angel of the jet is around the critical value $\theta_{\rm c}$ = arccos$\beta$ for the maximal apparent speed at a given $\beta$.
At this angle, the $\delta$ $\sim$ $\Gamma$ $\sim$ $\beta_{\rm app}$ = 15.33, and the intrinsic jet speed $\beta$ = $\sqrt{1-\frac{1}{\Gamma^{2}}}$ is about 0.998c. We can get the viewing angle $\theta_{\rm c}$  = 3.73$\degree$ $\pm$ 0.21$\degree$ and the intrinsic jet opening angle is $\rm \alpha_{int}$ = 1.78$\degree$ $\pm$ 0.4 $\degree$. With this small viewing angle, one can classify Fermi J1544-0649 as blazar.

Based on the above result, the viewing angle should be around 3.7$\degree$ to 7.4$\degree$. We suggest that the object is a possible misaligned blazar ($>$ 7$\degree$), but we cannot exclude the possibility of a blazar with a lower value of vlewing angle of 3.7$\degree$.

\subsection{Spectral index maps}

\begin{figure}
    \centering
    \includegraphics[width=0.45\textwidth]{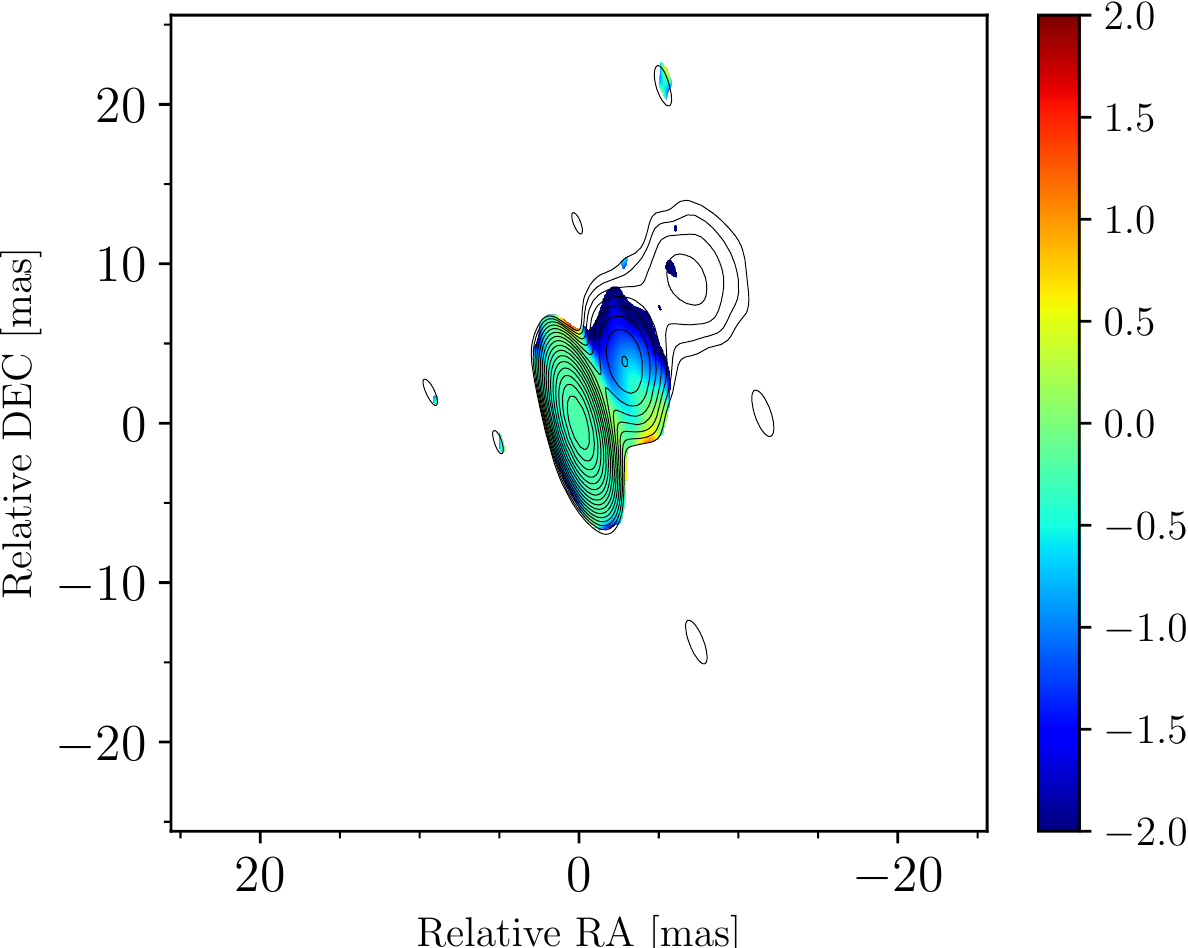}
    \caption{VLBA spectral index map od Fermi J1544-0649 from February 2019 data. The color scale represents the
spectral index distribution between 4.87 and 8.37 GHz, overlaid on the
contours showing the 4.87 GHz map of the source.  Peak intensity is
23.18 mJy beam$^{-1}$, the lowest contour level is at 0.24 mJy beam$^{-1}$ corresponding to 3$\sigma$ image noise.}
    \label{fig:spix}
\end{figure}

Since the 5 and 8 GHz observations were performed simultaneously with one day separation between the frequency bands, spectral information is available.
We obtained three spectral index maps using the VIMAP program \citep{2014JKAS...47..195K} from the 5- and 8-GHz VLBA data by aligning the maps on the core positions.
The 8-GHz images were imaged with the same pixel size (0.05 mas) and restoring beam size as the 5-GHz maps.
Accurate alignment of maps obtained at two different frequencies requires alignment of the optically thin jet at 5 and 8 GHz via spatial cross-correlation product.
Because of the relatively close frequencies, the shift is only $\sim$0.05 mas.
Spectral index maps obtained at three epochs do not show significant differences; accordingly, we only present one of them in Fig. \ref{fig:spix} as a representative example.

Our spectral index maps confirm the core–jet structure. 
Indeed the brightest southern component has flat spectral index, $\sim$0.1, as expected from the VLBI core of beamed AGN, while the northern region has a steeper spectrum with a spectral index of $\sim$ $-$ 1.0 typical for optically thin AGN jets (e.g., \citealt{2014AJ....147..143H}).

\section{multi-wavelength and multi-messenger observations}\label{mulobs}

\subsection{Swift-XRT/UVOT observations}\label{swiftdata}

The Neil Gehrels Swift Observatory (Swift) has observed the source 7 times in 2019. All XRT \citep{2005SSRv..120..165B}
observations were performed in photon counting mode.
Clean and calibrated files were obtained using the task xrtpipeline. We selected level 2 cleaned event files.

For Swift-UVOT data reduction, all extensions of sky images were stacked with uvotimsum. The UVOT observations have been taken in different filters and each
one analysed separately. The source magnitudes were computed using the uvotsource tool (HEASOFT v6.21)) with 3-$\sigma$ significance level.
The background was estimated from an annular region in the best source position with radii from 10$''$ to 20$''$.

\begin{figure*}
    \centering
    \includegraphics[width=0.7\textwidth]{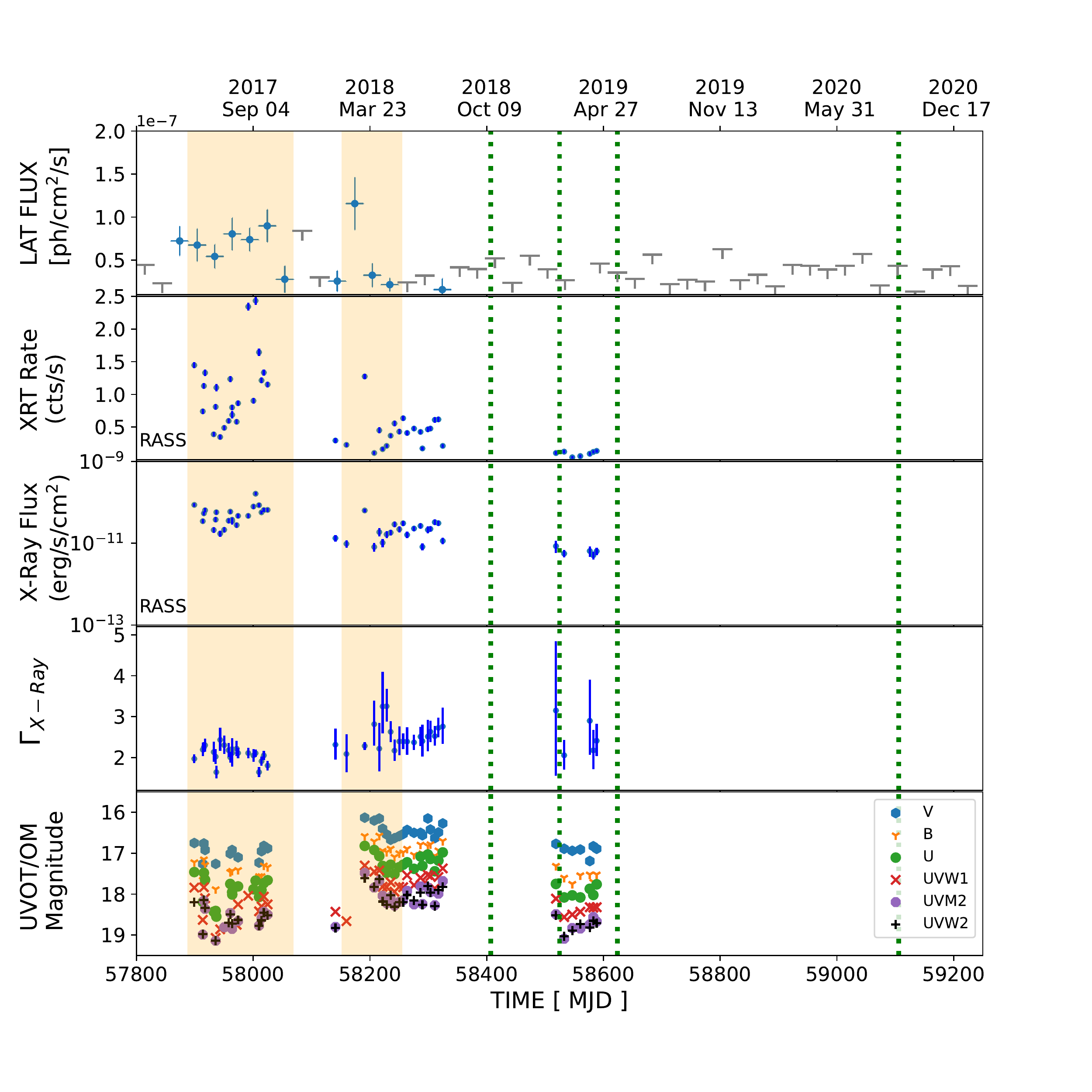}
    \caption{From top to bottom: The Fermi-LAT 0.1-100 GeV photon flux, Swift-XRT 0.3-10 keV count rate, energy flux, photon index, Swift-UVOT magnitudes of various filters between February 2017 and February 2021. The orange areas correspond to the time window of the gamma-ray flares in 2017 and 2018 respectively. The four green dotted lines correspond to the time of VLBA observations, which are observed in a quiescent period. Data prior to 2019 could be seen in \citealt{tam2020multi}. }
    \label{fig:mul_via}
\end{figure*}

The Swift-UVOT light curves (extinction not corrected) of different filters are plotted in Fig. \ref{fig:mul_via}. The X-ray flux goes to a lower state in 2019 compared to that prior to 2019. The UV emission was back to a low level of activity same with that in 2017.

\subsection{Fermi-LAT observations}\label{fermidata}

In our work, we use 4 years of Pass 8 (P8R3) Source class events collected between January 1, 2017 to January 25, 2021, and the fermi tools version 2.0.0 is used to reduce and analyze the Fermi-LAT data. Our region of interest (ROI) is 20$\degree$ $\times$ 20$\degree$ center at the position of 4FGL 1544.3-0649 (Ra=236.0785, Dec=-6.8255) whcih is associated with Fermi 1544-0649 in the newest 4FGL catalog.
We perform a binned maximum-likelihood analysis in our work. To reduce the contamination from Earth albedo $\gamma$-rays, events with zenith angles greater than 100$\degree$ were excluded. The diffuse model gll\_iem\_v07.fits (Galactic diffuse emission) and iso\_P8R3\_SOURCE\_V2\_v1.txt (isotropic diffuse component) are used to suppress the background, and sources in the Fermi catalog are included as background sources. We set free the spectral parameters of the sources within 5 $\degree$ from the ROI center (including the normalizations of the Galactic diffuse background and of the isotropic diffuse component) in each analysis. “FRONT+BACK” data are used to investigate the spectra, The spectrum of Fermi 1544-0649 is assumed to be a power-law(PL):

$$ \frac{dN}{dE} = N_0(\frac{E}{E_0})^{- \Gamma} $$

To check the GeV flare whether existed from 2017 to 2021, we plot the light curve with 30-day bins, the detection are defined as have a test statistic(TS) value $>$25, the upper limits are calculated at the 90$\%$ with the TS$<$9. the results are shown in \ref{fig:mul_via}. The flare from May 15, 2017 to Nov 11, 2017 correspond to a 14.02 $\sigma$ excess. The possible flare from Feb 1, 2018 to May 19, 2018 correspond to a 4.85 $\sigma$ excess. There is no GeV flare being detected after 2018.

\subsection{IceCube observations}\label{icedata}

In this search, we use 10 years data from 2008.04 to 2018.07 with an effective livetime of 3576.1 days \citep{2021arXiv210109836I}. This analysis selects track-like events because of their better angular resolution with a typical angular resolution of $\leq 1 \degree$.

66 neutrino events are taken in a circle with radius of 1 degree center at the position of 4FGL 1544-0649. We choose 24 of them by setting an energy threshold of 10 TeV to reduce the influence from background of atmospheric muons and neutrinos.

During the first three years, from Apr 6, 2008 to May 15, 2011, IceCube was incomplete and functioned with 40, 59, and 79 strings.
The data were taking of the full detector (IC86) since May 13, 2011, but the event selection and reconstruction was updated until it stabilized in 2012 \citep{PhysRevLett.124.051103}. For this reason, four event after 2012 are selected finally and listed in Table \ref{tab:icedata}.
We calculate a 3.1 $\sigma$ excess with the expected atmospheric neutrino background of 1 event for the period from 2012 to 2018. The number of background is calculated with effective area, livetime and atmospheric background flux of IceCube, with a solid angle of 0.001 sr. 

\begin{deluxetable}{ccccc}[!htp]
\centering
\tablenum{4}
\tablecaption{Selected neutrino events after 2012 from the position of Fermi J1544-0649.\label{tab:icedata}}
\tabletypesize{\small}
\tablewidth{0pt}
\tablehead{
MJD & RA (deg) & Dec (deg) & $\sigma$ (deg) & $\log_{10}(E/{\rm GeV})$ 
}

\startdata
56583 & 236.826 & -7.370 & 0.921 & 4.24 \\
56879 & 236.138 & -6.660 & 0.176 & 4.28 \\
57335 & 235.568 & -7.064 & 0.560 & 4.43 \\
57692 & 236.337 & -7.232 & 0.481 & 4.07 \\
\enddata
\end{deluxetable}

\section{discussion}

\subsection{the mystery GeV flare}\label{wobbling}

Fermi 1544-0649 remain quiet in GeV band until 2017 GeV/X-ray flares, after 2018, this source went back to a quiet period in GeV, X-ray and optical band. See fig \ref{fig:mul_via} , our radio observations lies in this quiet period.
Combined with the velocity of the jet, the most likely component be responsible for this burst is J4 in fig \ref{fig:pm}. We also can not rule out that other jet components buried in the core region may be responsible for the burst. Noticeably, the famous HST-1 of M87 experience sudden flare when this component was far from the core region.

Two models are usually used to explain such "misaligned blazar scenario". The former one refer to a theory that minijets caused by relativistic turbulence or magnetic reconnection beam their emission outside the jet cone \citep{Giannios:2009pi}, this model is usually proposed for the ultrafast flares($ \sim $ 1 day). The GeV flare in 2017 may consist of  unresolved fast flare similar to the $<$ 1 hour X-ray flare ; the latter one suggest that jet wobbling motion would result in a quasi-periodic multi-wavelength emission \citep{lico2020parsec}.
The explanation for this model is periodic variations of Doppler beaming factor produced by changes of the viewing angle from a geometrical origin. These changes in orientation may be related to the jet precessing or rotational motion, and/or helical structure within relativistic jets (\citealt{Rieger_2004,2015}).
If this model is true, the GeV flare has a more than 10 years duty cycle.

As mentioned above, the possible value of viewing angle could be as low as 3.7 $\degree$. In this scenario, Fermi J1544-0649 may experience a transition from FSRQs to the BL Lac
mode \citep{2002ApJ...571..226C}.
Apparently, after 2018, Fermi J1544-0649 show no blazar character. One may also argue that Fermi J1544-0649 has a very low duty cycle of gamma-ray emission \citep{10.1111/j.1365-2966.2004.08119.x}. 
\subsection{Is Fermi J1544-0649 a neutrino candidate?}

 Fermi J1544-0649 as a transient blazar, with a possible large viewing angle 7.42 degree, may be a high-energy neutrinos source. Though BL Lacs are normally considered with the low neutrino production efficiency based on the leptonic scenarios \citep{1985A&A...146..204G}, recent finding of the association of IceCube-170922A and TXS 0506+056, a BL Lac type blazar, gives another point of view. A 3.23 $\sigma$ post trial excess of HBLs and IBLs Blazars has also been reported \citep{10.1093/mnras/staa2082}. All these findings provide a stronger growing evidence for a connection between high-energy neutrinos and blazars.

The minijet/wobbling motion mentioned in section \ref{wobbling} is expected to throw out some material with a large inclination angle to the earth. As Tam 2020 found the BAL like feature, which indicates the clouds along our line of sight. If existed, it would increase the chance of the collision between jet and the clouds. The pp  or  p$\gamma$ interaction will happen and product pi-Mesons($\pi^0, \pi^{\pm} $), and $\pi^{\pm} $ will decay into lepton and corresponding neutrino.
For this reason, we searched ten years' archival neutrino data \citep{2021arXiv210109836I} from 2008 to 2018 and found 4 neutrino events listed in Table \ref{tab:icedata}. The four neutrino events were detected prior to the GeV/X-ray burst in 2017.

For the case of TXS 0506+056, IceCube Collaboration found a neutrino emission from this source in 2014-2015 which is statistically independent of the 2017 flaring episode in gamma rays and other wavelengths \citep{icecube2018neutrino}. It is considered that in neutrino production site, there is a significant attenuation of GeV gamma rays. Which further supposes a presence of enough dense X-ray target photons in the neutrino production region \citep{Inoue_2020}. If this is the case for certain fraction of high-energy sources, the neutrino event will shed the light on the weak X/gamma-ray emission source, not listed in the current catalogues. For sure, it needs more statistics to confirm, the follow-up search of neutrino event need to be performed, and a multi-wavelength search for such transient blazars need to be done.

\section{CONCLUSION}

In this article, we give the observed result of Fermi J1544-0649 using Very Long Baseline Interferometry (VLBI) data from 2018 to 2020. The position of the radio core is consistent with Gaia observation. The radio spectral index steepens along the jet, which confirm the core-jet structure. There are 4 jet components(J1, J2, J3, J4) being detected and their jet speeds are given respectively.  We derived the projected opening angle as $\rm \alpha_{pro}$ = 2arctan(D/2r) = 26.9$\degree$ $\pm$ 2.2 $\degree$.
Moreover, we derive the viewing angle of jet as 7.42 $\degree$ $\pm$ 1.68$\degree$ by using J3's speed, which is relatively large when compared to a sample of blazars seen by MOJAVE~\citep{2017MNRAS.468.4992P}.
This viewing angle lies on the boundary between the ones of blazars and the ones of galaxies, together with the low redshift, they seem to support the "misaligned blazar scenario". However, we can also derive a lower limit of viewing angle of 3.73 $\degree$, in such case,  Fermi J1544 may be a blazar with an extreme low duty cycle of gamma-ray emission.The multi-wavelength light curve of this source shows that Fermi 1544-0649 went to a quiet period in GeV, X-ray and optical band after 2018. During the radio observations time, no X/$\gamma$-ray flare was occured again. A doppler factor $\delta = 1.2 \pm 0.2$ during this non-flare time is given. Which is consistent with the "misaligned blazar scenario". Among the 4 components, J4 is most likely responsible for the 2017 GeV/X-ray flare.

As BL Lac could be a potential astrophysical neutrino source, we also analyse the data at the position of Fermi J1544-0649 from IceCube and find 4 neutrino events above 10 TeV with an excess of 3.1 $\sigma$ with the expected atmospheric neutrino background of 1 event for the searching period.

\section{ACKNOWLEDGEMENT}
This work is supported by the National Natural Science Foundation of China (NSFC) grants 11633007, 12005313 and U1731136, Guangdong Major Project of Basic and Applied Basic Research (Grant No. 2019B030302001), Key Laboratory of TianQin Project (Sun Yat-sen University), Ministry of Education, and the China Manned Space Project (No. CMS-CSST-2021-B09). 
X.-P. Cheng and B.-W. Sohn were supported by Brain Pool Program through the National Research Foundation of Korea (NRF) funded by the Ministry of Science and ICT (2019H1D3A1A01102564).
Y.Z. thanks NSFC grant number 11973099 for financial support. Z. L. Zhang thanks for support by National SKA Program of China (No. 2020SKA0120200). This work made use of the LAT data and Fermitools available at the Fermi Science Support Center (FSSC).
We thank the Help Desk of FSSC for their useful advice on our joint analyses with “PSF2” and “PSF3” data.
The VLBA observations were sponsored by Shanghai Astronomical Observatory through an MoU with the NRAO (Project code: BT146).

The Very Long Baseline Array is a facility of the National Science Foundation operated under cooperative agreement by Associated Universities, Inc. 
This work is made use of the East Asian VLBI Network (EAVN), which is operated under cooperative agreement by National Astronomical Observatory of Japan (NAOJ), Korea Astronomy and Space Science Institute (KASI), Shanghai Astronomical Observatory (SHAO), Xinjiang Astronomical Observatory (XAO), Yunnan Astronomical Observatory (YNAO), and National Geographic Information Institute (NGII), with the operational support by Ibaraki University (for the operation of HIT32 and TAK32), Yamaguchi University (for the operation of YAM32), and Kagoshima University (for the operation of VERA Iriki antenna).

\vspace{5mm}
\facilities{HST(STIS), Swift(XRT and UVOT), AAVSO, CTIO:1.3m,
CTIO:1.5m,CXO, VLBA, EAVN}

\software{
{\tt AIPS} \citep{2003ASSL..285..109G}, 
{\tt Astropy} \citep{Astropy}, 
{\tt DIFMAP} \citep{1994BAAS...26..987S}
          }

\bibliography{fermi1544}
\bibliographystyle{aasjournal}

\end{document}